\documentclass[english,preprint,nofootinbib]{revtex4}%
\usepackage[colorlinks=true,urlcolor=black,linkcolor=black,citecolor=black]%
{hyperref}
\usepackage[T1]{fontenc}
\usepackage[latin9]{inputenc}
\usepackage{amsmath}
\usepackage{amssymb}
\usepackage{amsfonts}
\usepackage{babel}
\usepackage{graphicx}%
\providecommand{\U}[1]{\protect\rule{.1in}{.1in}}
\makeatletter
\@ifundefined{textcolor}{}
{
\definecolor{BLACK}{gray}{0}
 \definecolor{WHITE}{gray}{1}
 \definecolor{RED}{rgb}{1,0,0}
 \definecolor{GREEN}{rgb}{0,1,0}
 \definecolor{BLUE}{rgb}{0,0,1}
 \definecolor{CYAN}{cmyk}{1,0,0,0}
 \definecolor{MAGENTA}{cmyk}{0,1,0,0}
 \definecolor{YELLOW}{cmyk}{0,0,1,0}
 }
\makeatother
\begin{document}
\title{Asymptotically locally flat spacetimes \\and dynamical black flowers in three dimensions}
\author{Glenn Barnich$^{1}$, Cédric Troessaert$^{2}$, David Tempo$^{1,2}$, Ricardo
Troncoso$^{2}$}
\email{gbarnich@ulb.ac.be, troessaert@cecs.cl, tempo@cecs.cl, troncoso@cecs.cl}
\affiliation{$^{1}$Physique Théorique et Mathématique, Université Libre de Bruxelles and
International Solvay Institutes, Campus Plaine C.P. 231, B-1050 Bruxelles,
Belgium, }
\affiliation{$^{2}$Centro de Estudios Científicos (CECs), Av. Arturo Prat 514, Valdivia, Chile.}
\preprint{CECS-PHY-15/10}

\begin{abstract}
The theory of massive gravity proposed by Bergshoeff, Hohm and Townsend is
considered in the special case of the pure irreducibly fourth order quadratic
Lagrangian. It is shown that the asymptotically locally flat black holes of
this theory can be consistently deformed to \textquotedblleft black
flowers\textquotedblright\ that are no longer spherically symmetric. Moreover,
we construct radiating spacetimes settling down to these black flowers in the
far future. The
generic case can be shown to fit within a relaxed set of asymptotic conditions
as compared to the ones of general relativity at null infinity, while the
asymptotic symmetries remain the same. Conserved charges as surface integrals
at null infinity are constructed following a covariant approach, and their
algebra represents BMS$_{3}$, but without central extensions. For solutions
possessing an event horizon,
we derive the first law of thermodynamics from these surface integrals.

\end{abstract}
\maketitle

\section{Introduction}

In three spacetime dimensions, since the Riemann tensor can be expressed in
terms of the Ricci tensor and the scalar curvature, general relativity in
vacuum does not admit propagating degrees of freedom. Furthermore, black hole
solutions in vacuum exist only in the case of negative cosmological constant
\cite{BTZ,BHTZ}. Thus, it is interesting to explore whether different simple
models of three-dimensional gravity could capture further key properties of
gravity in four dimensions. This seems to be the case for the theory of
massive gravity proposed by Bergshoeff, Hohm and Townsend (BHT) \cite{BHT0,BHT1}.
Hereafter, we consider this theory with the purely quadratic action given by
\begin{equation}
I\left[  g\right]  =\frac{1}{16\pi G}\int d^{3}x\sqrt{-g}\left(  R_{\mu\nu
}R^{\mu\nu}-\frac{3}{8}R^{2}\right)  \ .\label{ActionK}%
\end{equation}
The corresponding field equations
\begin{equation}
2\square R_{\mu\nu}-\frac{1}{2}\nabla_{\mu}\nabla_{\nu}R-\frac{1}{2}\square
Rg_{\mu\nu}+4R_{\mu\sigma\nu\rho}R^{\sigma\rho}-\frac{3}{2}RR_{\mu\nu
}-R_{\sigma\rho}R^{\sigma\rho}g_{\mu\nu}+\frac{3}{8}R^{2}g_{\mu\nu
}=0\;,\label{eq:FEq}%
\end{equation}
are irreducibly of fourth order and propagate a single degree of freedom
\cite{BHT1,Deser-Alas}. Remarkably, these field equations admit a static
asymptotically locally flat black hole solution, whose metric reads
\cite{OTT}
\begin{equation}
ds^{2}=-\left(  br-\mu\right)  dt^{2}+\frac{dr^{2}}{br-\mu}+r^{2}d\phi
^{2}\ .\label{BH-metric}%
\end{equation}
This spacetime is conformally flat\footnote{Indeed, this class of black holes
was first found for conformal gravity in vacuum \cite{Joao-Pessoa}, whose
field equations imply the vanishing of the Cotton tensor. For this reason, it has been shown that these black holes also solve the field equations of the Poincar\'e  gauge theory 
\cite{Blagojevic}.}, and its Ricci
scalar is given by
\begin{equation}
R=-\frac{2b}{r}\ ,
\end{equation}
so that it possesses a spacelike singularity at the origin, which is
surrounded by an event horizon located at $r_{+}=\mu/b$, provided $b>0$ and
$\mu>0$. In the case of $\mu=0$, there is a NUT on top of the null
singularity. Note that its Hawking temperature, which turns out to be
\begin{equation}
T=\frac{b}{4\pi}\ ,\label{Temperature}%
\end{equation}
depends only on one of the integration constants in (\ref{BH-metric}). One of
our purposes is to compute the mass of this black hole from a surface
integral, which is not a simple task for a fourth-order theory with quadratic
terms in the curvature. 

In the next section, we show how the black hole solutions described by
(\ref{BH-metric}) can be deformed to \textquotedblleft black
flowers\textquotedblright\ that are no longer spherically symmetric. Their
extension to time-dependent Robinson-Trautman-like solutions is also presented. In section
\ref{Asymptotic-Conditions-Symmetries}, the asymptotic conditions of general relativity at null infinity
\cite{ABS-AS-Null-3D-GR,GC-Null-Conditions-3D,BMSaspect-1}
are relaxed to accommodate these solutions and the
asymptotic symmetries are worked out. In Section \ref{Surface-Charges},
conserved charges as surface integrals at null infinity are constructed
following a covariant approach
\cite{Covariant-Barnich-Brandt,GC-Boundary-Charges}. Their algebra corresponds
to BMS$_{3}$ without central extensions. The global charges of the (rotating)
black holes as well as the ones of the dynamical black flowers are explicitly
computed in Section \ref{Surface-Charges-RotBHs-BFlowers}. Section
\ref{Thermodynamics} is devoted to the derivation of the first law of
thermodynamics from the surface integrals when event horizons
are present.

\section{Dynamical black flowers}

It is useful to express the black hole metric (\ref{BH-metric}) in terms of a
set of coordinates adapted to null infinity. This can easily be done
by setting $u=t-r^{\ast}$, where $r^{\ast}$ stands for the tortoise coordinate
defined through $dr^{\ast}=\frac{dr}{br-\mu}$, so that the black hole metric
read
\begin{equation}
ds^{2}=-\left(  br-\mu\right)  du^{2}-2dudr+r^{2}d\phi^{2}%
\ .\label{StaticMetric}%
\end{equation}
These metrics can be deformed along the spacelike Killing vector
$\partial_{\phi}$, yielding
\begin{equation}
ds^{2}=-\left(  br-\mu\right)  du^{2}-2dudr+\left(  r-\mathcal{H}\left(
u,\phi\right)  \right)  ^{2}d\phi^{2}\;,\label{eq:AnzatDBF}%
\end{equation}
where $\mathcal{H}\left(  u,\phi\right)  $ is periodic in the angular
coordinate $\phi$. Even though the deformation is not of Kerr-Schild type, the
field equations (\ref{eq:FEq}) surprisingly reduce to a single linear
differential equation given by
\begin{equation}
\partial_{u}\left(  \partial_{u}\mathcal{H}+\frac{b}{2}\mathcal{H}\right)
=0\;.\label{eq:LDE}%
\end{equation}
The general solution reads
\begin{equation}
\mathcal{H}\left(  u,\phi\right)  =\mathcal{A}\left(  \phi\right)
+\mathcal{B}\left(  \phi\right)  e^{-\frac{b}{2}u}\;,\label{eq:H-DBF}%
\end{equation}
where $\mathcal{A}\left(  \phi\right)  $ and $\mathcal{B}\left(  \phi\right)
$ are arbitrary periodic functions. Note that the deformed metrics are still
conformally flat. Their Ricci scalar is%
\begin{equation}
R=\frac{2b}{r-\mathcal{H}}\;,
\end{equation}
signaling the existence of a curvature singularity at 
$r=\mathcal{H}(u,\phi)$, so that the range of the radial coordinate can be
chosen as $\mathcal{H}<r<\infty$.

This class of solutions can be divided as follows: when $\mathcal{B}\left(
\phi\right)  =0$, the resulting metrics are static while $\mathcal{B}\left(
\phi\right)  \neq0$ leads to metrics describing dynamical spacetimes.\newline

In the static case, the metric has an
event horizon located at $r_{+}=\mu/b$ . In order for this horizon to enclose
the curvature singularity, the function $\mathcal{A}\left(
\phi\right)  $ has to be bounded from above according to%
\begin{equation}
\mathcal{A}\left(  \phi\right)  <r_{+}\ .\label{Bound}%
\end{equation}
The induced metric on a
constant $u$ slice of the horizon is given by
\begin{equation}
ds_{h}^{2}=\left(  r_{+}-\mathcal{A}\left(  \phi\right)  \right)  ^{2}%
d\phi^{2}\;,
\end{equation}
and since the function $\mathcal{A}\left(  \phi\right)  $ is arbitrary, the
horizon is generically not spherically symmetric. In particular, these
spacetimes can
describe a black flower. The associated temperature can be
evaluated using the surface gravity and is found to coincide with the one of
the spherically symmetric black holes given in (\ref{Temperature}).

In the dynamical case, the deformation
$\mathcal{H}\left(  u,\phi\right)  $ propagates along outgoing null rays. At
late retarded time, $u$ going to infinity, this generic configuration tends to
a static black flower. Therefore, one may view these solutions as describing a
black flower surrounded by an outgoing graviton. This can be easily seen
introducing Kruskal-Szekeres coordinates
\begin{equation}
U=-e^{-\frac{b}{2}u}\;;\;V=e^{\frac{b}{2}v}\;;\;r\left(  U,V\right)  =\frac
{1}{b}\left(  \mu-UV\right)
\end{equation}
where $v=t+r^{\ast}$ is the advanced time. The metric then takes the form
\begin{equation}
ds^{2}=-\frac{4}{b^{2}}dUdV+\left(  r\left(  U,V\right)  -\mathcal{H}\left(
U,V,\phi\right)  \right)  ^{2}d\phi^{2}\;,
\end{equation}
with
\begin{equation}
\mathcal{H}\left(  U,V,\phi\right)  =\mathcal{A}\left(  \phi\right)
-\mathcal{B}\left(  \phi\right)  U\;.
\end{equation}
This indicates that the spacetime possesses a horizon at $U=0$
with a curvature singularity located at $r\left(  U,V\right)  =\mathcal{H}%
\left(  U,V,\phi\right)  $, i.e.
\begin{equation}
U\left(  V-b\mathcal{B}\left(  \phi\right)  \right)  =\mu-b\mathcal{A}\left(
\phi\right)  \;.
\end{equation}
The future curvature singularity is surrounded by the horizon only when
eq. (\ref{Bound}) holds. On this horizon the metric reduces to the one of the
static black flower while outside of the horizon, for $U<0$, the spacetime is
filled with outgoing radiation.

These solutions are similar to the Robinson-Trautman spacetimes found in 
four-dimensional Einstein gravity \cite{RobinsonTrautman-1,RobinsonTrautman-2},
which describe Schwarzschild black holes surrounded by outgoing radiation. The
main difference is that the time evolution of these four-dimensional analogues
is described by a complicated fourth order nonlinear differential equation.

It is worth mentioning that deformations of exact solutions
that are not of the standard Kerr-Schild type (see e.g.
\cite{KS-Extended-1,KS-Extended-2}), have been found for different cases
of BHT massive gravity theory \cite{KC-Anzats-BHT-1}, and also for
Einstein-Gauss-Bonnet \cite{KC-Anzats-EGB-1, KC-Anzats-EGB-2}\ and Lovelock
theories in higher dimensions \cite{KC-Anzats-Lovelock-3}.

\section{Relaxed fall-off conditions and asymptotic symmetries}

\label{Asymptotic-Conditions-Symmetries}  Recently, there has been
renewed interest in  asymptotically flat spacetimes at null infinity
in three and four dimensions (see, e.g., \cite{BMSaspect-1,
  Barnich:2006av,Barnich:2012aw,Bagchi:2012yk,CosmoEntro-1,FlatHolography-1,%
Bagchi:2013lma,Barnich:2013axa,stromingerBMS-2,%
stromingerBMS-1,StromingerBMS-3,Strominger:2014pwa,Barnich:2015mui}.
Since the dynamical black flowers contain outgoing gravitational
radiation, it is natural to study their asymptotic structure at null
infinity.  The metric (\ref{eq:AnzatDBF}) describing these
dynamical black flowers can be shown to fit within a set of
asymptotic conditions that are relaxed as compared to the standard
ones for general relativity
\cite{ABS-AS-Null-3D-GR,GC-Null-Conditions-3D}. Nevertheless, the
asymptotic symmetry algebra remains the same.  The suitably
relaxed asymptotic conditions for the metric are
\begin{equation}
\Delta g_{AB}=h_{AB}r+\mathcal{O}\left(  1\right)  \;;\;\Delta g_{rr}%
=\mathcal{O}\left(  r^{-3}\right)  \;;\;\Delta g_{Ar}=\mathcal{O}\left(
r^{-1}\right)  \;,\label{BoundaryConditions}%
\end{equation}
where the functions $h_{AB}$ depend only on $x^{A}=\left(  u,\phi\right)  $
and $\Delta g_{\mu\nu}=g_{\mu\nu}-\bar{g}_{\mu\nu}$ correspond to 
deviations from the Minkowski metric,
\begin{equation}
d\bar{s}^{2}=-du^{2}-2dudr+r^{2}d\phi^{2}\ .\label{Minkowski}%
\end{equation}
The asymptotic conditions (\ref{BoundaryConditions}) are mapped into
themselves under diffeomorphisms of the form,
\begin{align}
\xi^{u} &  =T\left(  \phi\right)  +u\partial_{\phi}Y\left(  \phi\right)
+\mathcal{O}\left(  r^{-1}\right)  \;,\nonumber\\
\xi^{r} &  =-r\partial_{\phi}Y\left(  \phi\right)  +\mathcal{O}\left(
1\right)  \;,\label{AsymptoticSymmetries}\\
\xi^{\phi} &  =Y\left(  \phi\right)  -\frac{1}{r}\partial_{\phi}\left(
T\left(  \phi\right)  +u\partial_{\phi}Y\left(  \phi\right)  \right)
+\mathcal{O}\left(  r^{-2}\right)  \ .\nonumber
\end{align}
The sub-space of vectors with vanishing $T(\phi),Y(\phi)$ form an ideal and
the quotient algebra is BMS$_{3}$, i.e., the semi-direct sum of the
vector fields on the circle
 with the abelian ideal of supertranslations. Thus, remarkably, the
relaxed behavior of the metric at infinity does not spoil the asymptotic
symmetries obtained in Refs.~\cite{ABS-AS-Null-3D-GR,GC-Null-Conditions-3D}
for general relativity in three dimensions.

\section{Conserved charges at null infinity}

\label{Surface-Charges}

In order to obtain the conserved charges associated to the asymptotic
symmetries spanned by (\ref{AsymptoticSymmetries}), it is useful to write the
action (\ref{ActionK}) in terms of an auxiliary field $\mathcal{\ell}_{\mu
\nu}$, so that the field equations reduce from fourth to second order
\cite{BHT0,BHT1,Deser-Alas}. The action then reads
\begin{equation}
I\left[  g,\mathcal{\ell}\right]  =\frac{1}{16\pi G}\int d^{3}x\sqrt
{-g}\left(  \mathcal{\ell}^{\mu\nu}G_{\mu\nu}-\frac{1}{4}\left(
\mathcal{\ell}^{\mu\nu}\mathcal{\ell}_{\mu\nu}-\mathcal{\ell}^{2}\right)
\right)  \ ,
\end{equation}
and the algebraic field equations associated to $\mathcal{\ell}_{\mu\nu}$ are
given by
\begin{equation}
G^{\mu\nu}-\frac{1}{2}\left(  \mathcal{\ell}^{\mu\nu}-g^{\mu\nu}\mathcal{\ell
}\right)  =0\ .
\end{equation}
On-shell, the auxiliary field turns out to be proportional to the Schouten tensor,
\begin{equation}
\mathcal{\ell}_{\mu\nu}=2\left(  R_{\mu\nu}-\frac{1}{4}g_{\mu\nu}R\right)
\ .\label{Lmunu}%
\end{equation}
Varying the action with respect to the metric also produces second order field
equations
\begin{multline}
\nabla^{\alpha}\nabla_{\alpha}\mathcal{\ell}^{\mu\nu}-2\nabla_{\rho}%
\nabla^{(\mu}\mathcal{\ell}^{\nu)\rho}+\nabla^{\mu}\nabla^{\nu}\mathcal{\ell
}+g^{\mu\nu}\left(  \nabla_{\rho}\nabla_{\lambda}\mathcal{\ell}^{\rho\lambda
}-\nabla^{\alpha}\nabla_{\alpha}\mathcal{\ell}\right)  +4\mathcal{\ell
}^{\lambda(\mu}G_{\hspace{0.05in}\lambda}^{\nu)}\\
+\mathcal{\ell}^{\mu\nu}R-\mathcal{\ell}R^{\mu\nu}-g^{\mu\nu}\mathcal{\ell
}^{\rho\sigma}G_{\rho\sigma}+\mathcal{\ell}^{\lambda\mu}\mathcal{\ell
}_{\hspace{0.05in}\lambda}^{\nu}-\mathcal{\ell\ell}^{\mu\nu}-\frac{1}{4}%
g^{\mu\nu}\left(  \mathcal{\ell}_{\alpha\beta}\mathcal{\ell}^{\alpha\beta
}-\mathcal{\ell}^{2}\right)  =0\;.
\end{multline}

\bigskip{}  Following the covariant approach described in
\cite{Covariant-Barnich-Brandt,GC-Boundary-Charges}, the conserved charges are
given by,
\begin{equation}
Q_{\xi}=\int_{0}^{1}ds\left( \frac{1}{2}\int_{\partial\Sigma}\varepsilon
_{\nu\mu\alpha}\tilde{k}_{\xi}^{\left[ \nu\mu\right] }dx^{\alpha}\right)
\ .\label{Charges}%
\end{equation}
In our case, the superpotential acquires the form
\begin{align}
\left( \frac{8\pi G}{\sqrt{-g}}\right) \tilde{k}_{\xi}^{\left[ \nu\mu\right]
}  &  =P^{\mu\rho\left( \lambda\nu\right) \alpha\beta}\left( \frac{2}{3}%
\xi_{\rho}\nabla_{\lambda}k_{\alpha\beta}-\frac{1}{3}k_{\alpha\beta}%
\nabla_{\lambda}\xi_{\rho}\right) -T^{\mu\rho\lambda\nu\alpha\beta\delta
\gamma}\left( \frac{1}{3}\xi_{\rho}h_{\alpha\beta}\nabla_{\lambda
}\mathcal{\ell}_{\delta\gamma}\right) \ \nonumber\\
&  +U^{\mu\rho(\lambda\nu)\alpha\beta\delta\gamma}\left( \frac{2}{3}\xi_{\rho
}\nabla_{\lambda}h_{\alpha\beta}-\frac{1}{3}h_{\alpha\beta}\nabla_{\lambda}%
\xi_{\rho}\right) \ell_{\delta\gamma}-(\mu\leftrightarrow\nu
)\ ,\label{Superpotential}%
\end{align}
with
\begin{align}
P^{\mu\nu\kappa\sigma\alpha\beta}  &  =g^{\kappa\sigma}g^{\mu\lbrack(\alpha
|}g^{\nu]|\beta)}+g^{\nu\sigma}g^{\mu\lbrack\kappa}g^{(\alpha]\beta
)}+g^{\kappa(\alpha|}g^{\mu\lbrack\nu}g^{\sigma]|\beta)}\ ,\\
U^{\mu\rho(\lambda\nu)\alpha\beta\delta\gamma}  &  =\hat{Y}_{\eta\varepsilon
}^{\left( \lambda\nu\right) \alpha\beta}P_{_{_{II}}}^{\mu\rho\gamma\delta
\eta\varepsilon}-2P^{\mu\rho(\lambda|\eta\varepsilon\gamma}\hat{X}%
_{\eta\varepsilon}^{\delta|\nu)\alpha\beta}+P^{\mu\delta(\lambda\nu
)\alpha\beta}\bar{g}^{\gamma\rho}\ ,\\
T^{\mu\rho\lambda\nu\alpha\beta\delta\gamma}  &  =U^{\mu\rho(\lambda\nu
)\alpha\beta\delta\gamma}+\frac{3}{2}\left( P^{\mu\rho\varepsilon\sigma
\gamma\delta}\hat{X}_{\sigma\varepsilon}^{\lambda\nu\alpha\beta}+2\left(
P^{\mu\rho\sigma\lambda\varepsilon\gamma}+P^{\mu\rho\lambda\sigma
\varepsilon\gamma}\right) \hat{X}_{\sigma\varepsilon}^{\delta\nu\alpha\beta
}\right) \ ,
\end{align}
and
\begin{align}
P_{_{_{II}}}^{\mu\nu\delta\lambda\alpha\beta}  &  =\left( g^{\mu(\lambda
}g^{\delta)(\beta|}-\frac{1}{2}g^{\delta\lambda}g^{\mu(\beta|}\right)
g^{|\alpha)\nu}+\left( g^{\mu(\alpha}\bar{g}^{\beta)(\delta|}-\frac{1}%
{2}g^{\alpha\beta}g^{\mu(\delta|}\right) g^{|\lambda)\nu}\nonumber\\
&  -\frac{1}{2}g^{\mu\nu}\left( g^{(\delta(\alpha}g^{\beta)\lambda)}-\frac
{1}{2}g^{\delta\lambda}g^{\alpha\beta}\right) \ ,
\end{align}
\begin{align}
\hat{Y}_{\eta\varepsilon}^{\kappa\sigma\alpha\beta}  &  =\frac{1}{2}\left(
-\delta_{\hspace{0.05in}(\eta}^{\alpha}\delta_{\hspace{0.05in}\varepsilon
)}^{\beta}g^{\sigma\kappa}-\delta_{\hspace{0.05in}(\eta}^{\kappa}%
\delta_{\hspace{0.05in}\varepsilon)}^{\sigma}g^{\alpha\beta}+2\delta
_{\hspace{0.05in}(\eta}^{\sigma}\delta_{\hspace{0.05in}\varepsilon)}^{(\alpha
}g^{\beta)\kappa}\right) \ ,\\
\hat{X}_{\sigma\beta}^{\lambda\eta\gamma\rho}  &  =\frac{1}{2}\bar{g}%
^{\lambda\xi}\left( 2\delta_{\hspace{0.05in}(\sigma}^{\eta}\delta
_{\hspace{0.05in}\beta)}^{(\gamma}\delta_{\hspace{0.05in}\xi}^{\rho)}%
-\delta_{\hspace{0.05in}\xi}^{\eta}\delta_{\hspace{0.05in}\sigma}^{(\gamma
}\delta_{\hspace{0.05in}\beta}^{\rho)}\right) \ .
\end{align}
Here, it is assumed that $g_{\mu\nu}:=$ $g_{\mu\nu}^{s}$ and $\ell_{\mu\nu}:=$
$\ell_{\mu\nu}^{s}$ are one-parameter families of solutions, interpolating
between the background fields $\bar{g}_{\mu\nu}=g_{\mu\nu}^{0}$, $\bar{\ell
}_{\mu\nu}=$ $\ell_{\mu\nu}^{0}$ and some given solution $g_{\mu\nu}=g_{\mu
\nu}^{1}$ and $\ell_{\mu\nu}=$ $\ell_{\mu\nu}^{1}$. The deviations $h_{\mu\nu
}$ and $k_{\mu\nu}$ are defined as the tangent vectors to $g_{\mu\nu}^{s}$ and
$\ell_{\mu\nu}^{s}$ in the solution space, i.e.,
\begin{align}
h_{\mu\nu}  &  =\frac{d}{ds}g_{\mu\nu}^{s}\;;\;k_{\mu\nu}=\frac{d}{ds}%
\ell_{\mu\nu}^{s}\text{ }.
\end{align}

\bigskip{}

As a first check, for the static black hole (\ref{StaticMetric}) the
interpolating metric $g_{\mu\nu}^{s}$ can be chosen as
\begin{equation}
g_{\mu\nu}^{s}\rightarrow ds_{s}^{2}=-(1+s\left(  br-\mu-1\right)
)du^{2}-2dudr+r^{2}d\phi^{2}\ ,\label{gs}%
\end{equation}
and $\ell_{\mu\nu}^{s}$ is obtained plugging (\ref{gs}) in eq. (\ref{Lmunu}).
In particular, for the mass, the surface integral gives
\begin{equation}
M=Q\left(  \partial_{u}\right)  =\frac{b^{2}}{32G}\ .\label{MQ1}%
\end{equation}

\bigskip Using an arbitrary set of interpolating metrics satisfying the
asymptotic conditions given in eq. (\ref{BoundaryConditions}), the conserved
charges (\ref{Charges}) simplify to
\begin{equation}
Q_{\xi}=Q\left[  T,Y\right]  =\frac{1}{64\pi G}\int_0^{2\pi} d\phi\left(  \left(
T+u\partial_{\phi}Y\right)  h_{uu}^{2}+2Yh_{u\phi}h_{uu}+4h_{uu}\partial
_{\phi}Y+4Y\partial_{u}h_{u\phi}\right)  \ .\label{Charges-Asympt}%
\end{equation}
The black hole mass in (\ref{MQ1}) can then also directly be computed from
(\ref{Charges-Asympt}), by taking into account that for the metric
(\ref{StaticMetric}), the only nonvanishing deviation from the background is
given by $h_{uu}=-b$.\bigskip{}

Using the leading terms of the following asymptotic field equations
\begin{align}
E_{uu} &  =-\frac{1}{4}\left(  \partial_{u}h_{uu}^{2}\right)  r^{-1}%
+\mathcal{O}\left(  r^{-2}\right)  =0\ ,\\
E_{u\phi} &  =\frac{1}{4}\left(  \partial_{\phi}h_{uu}^{2}-2\partial
_{u}\left(  h_{uu}h_{u\phi}\right)  +4\partial_{u}\left(  \partial_{\phi
}h_{uu}-\partial_{u}h_{u\phi}\right)  \right)  r^{-1}+\mathcal{O}\left(
r^{-2}\right)  =0\ ,
\end{align}
and  the action of the asymptotic symmetries on the relevant dynamical
fields
\begin{align}
\delta_{\xi}h_{uu} &  =\left(  T\left(  \phi\right)  +u\partial_{\phi}Y\left(
\phi\right)  \right)  \partial_{u}h_{uu}+\partial_{\phi}\left(  h_{uu}%
Y\right)  \;,\\
\delta_{\xi}h_{u\phi} &  =h_{uu}\partial_{\phi}\left(  T\left(  \phi\right)
+u\partial_{\phi}Y\left(  \phi\right)  \right)  +\left(  T\left(  \phi\right)
+u\partial_{\phi}Y\left(  \phi\right)  \right)  \partial_{u}h_{u\phi}%
+\partial_{\phi}\left(  h_{u\phi}Y\right)  \;,
\end{align}
it is simple to verify that the algebra $\left\{  Q_{\xi_{1}},Q_{\xi_{2}%
}\right\}  \equiv\delta_{\xi_{2}}Q_{\xi_{1}}$ realises the BMS$_3$ without central extension,
\begin{equation}
\left\{  Q_{\xi_{1}},Q_{\xi_{2}}\right\}  \approx Q_{\left[  \xi_{1},\xi_{2}\right]}.
\end{equation}
Expanding the charges in Fourier modes
\begin{equation}
\mathcal{P}_{n}=Q\left[  e^{in\phi},0\right]  \;;\;\;\mathcal{J}_{n}=Q\left[
0,e^{in\phi}\right]  \;,
\end{equation}
the algebra takes the familiar form
\begin{align}
i\left\{  \mathcal{J}_{m},\mathcal{J}_{n}\right\}   &  =\left(  m-n\right)
\mathcal{J}_{m+n}\;,\nonumber\\
i\left\{  \mathcal{J}_{m},\mathcal{P}_{n}\right\}   &  =\left(  m-n\right)
\mathcal{P}_{m+n}\;,\label{BMS3}\\
i\left\{  \mathcal{P}_{m},\mathcal{P}_{n}\right\}   &  =0\;.\nonumber
\end{align}

The vanishing of the potential central extensions of the algebra (\ref{BMS3})
can also be obtained by taking the flat limit of the
 extended two-dimensional conformal algebra given in
\cite{OTT}, and also in \cite{PTT,FareghbalHosseini} for BHT massive gravity in the
special case (with a unique maximally symmetric vacuum) from a holographic approach.

\section{conserved charges for rotating black holes and dynamical black
flowers}

\label{Surface-Charges-RotBHs-BFlowers}

The theory under consideration also admits a rotating asymptotically locally
flat black hole solution \cite{OTT,GOTT,PTT}. In outgoing null coordinates,
the metric takes the form
\begin{equation}
ds^{2}=-NFdu^{2}-2N^{\frac{1}{2}}dudr+\left(  r^{2}+r_{0}^{2}\right)  \left(
d\phi+N^{\phi}du\right)  ^{2}\ ,\label{RotatingBH}%
\end{equation}
where
\[
F=br-\mu\text{\ };\text{\ }N=\frac{\left(  8r+a^{2}b\right)  ^{2}}{64\left(
r^{2}+r_{0}^{2}\right)  }\text{\ };\text{\ }N^{\phi}=-\frac{a}{2}\left(
\frac{br-\mu}{r^{2}+r_{0}^{2}}\right)  \text{\ };\text{\ }r_{0}^{2}%
=\frac{a^{2}}{4}\left(  \mu+\frac{a^{2}b^{2}}{16}\right)  \text{ }.
\]
Note that when the angular parameter $a$ vanishes, the solution reduces to the
static black hole given in (\ref{StaticMetric}). Since the relevant deviations
from the background are given by $h_{uu}=-b$ and $h_{u\phi}=-\frac{a}{2}b$,
the mass and the angular momentum can be directly obtained using
(\ref{Charges-Asympt}). Thus, one finds
\begin{align}
M=Q\left(  \partial_{u}\right)   &  =\frac{b^{2}}{32G}\ ,\\
J=Q\left(  \partial_{\phi}\right)   &  =\left(  \frac{b^{2}}{32G}\right)
a=Ma\;,
\end{align}
which is in agreement with the vanishing cosmological constant limit of BHT
massive gravity in the special case that admits a unique maximally symmetric
vacuum \cite{PTT,FareghbalHosseini}. Furthermore, one deduces that the only
nonvanishing global charges associated with the rotating black hole are the
mass and the angular momentum, and therefore, the integration constant $\mu$
turns out to be a \textquotedblleft gravitational hair\textquotedblright%
\ parameter\footnote{Curiously, if one compares these results with the ones in
\cite{OTT, GOTT, PTT} in the presence of cosmological constant, the roles that
the integration constants $b$ and $\mu$ play in the global charges become
interchanged.}.

In the case of the dynamical black flowers (\ref{eq:AnzatDBF}), (\ref{eq:H-DBF}%
), the deformation with respect to the static black hole does not modify the
values of $h_{uu}$ and $h_{u\phi}$. Therefore, the charges
remain the same, that is $M=\frac{b^{2}}{32G}$ and $J=0$. Surprisingly, even
though we have outgoing radiation the total mass at null infinity is constant,
i.e., there is no \textquotedblleft news\textquotedblright%
\ \cite{Bondi-News-1,Sachs-News-2}. When restricting to static black
flowers, it is worth noticing that both $\mu$ and $\mathcal{A}\left(
\phi\right)  $ are hair parameters. In this sense, from the mode expansion of
$\mathcal{A}\left(  \phi\right)  $, one might interpret static black
flowers as black hole solutions with an infinite number of purely
gravitational hair parameters.

\section{Thermodynamics}

\label{Thermodynamics}

The conserved charges (\ref{Charges}) are also useful in the presence
of horizons. Indeed, for a black hole with a Killing horizon, as for
the static black flowers, one can use surface charges for
$\xi=\partial_{u}$ to derive thermodynamical properties of the black
holes as in \cite{Wald-Entropy,Iyer:1994ys}.

The conservation of the superpotential $\tilde{k}_{\xi}^{[\mu\nu]}$ associated
to a Killing vector $\xi$ 
implies
\begin{equation}
\delta\left(  \left.  Q_{\xi}\right\vert _{r=\infty}+\left.  Q_{\xi
}\right\vert _{r=r_{+}}\right)  =0\ .\label{cons}%
\end{equation}
Remarkably, for the static black flowers the superpotential can be computed
exactly, and is simply given by
\begin{equation}
\tilde{k}_{\xi}^{\left[  ur\right]  }=\frac{b\delta b}{32\pi G}\;.
\end{equation}
As seen above, evaluating the charge at infinity one obtains the total energy,
i.e.,
\begin{equation}
\left.  Q_{\xi}\right\vert _{r=\infty}=\frac{b^{2}}{32G}=M\ ,
\end{equation}
while the variation of the surface integral at the horizon $r=r_{+}$ gives
\begin{equation}
\left.  \delta Q_{\xi}\right\vert _{r=r_{+}}=-\frac{b\delta b}{16G}=-T\delta
S\ .
\end{equation}
The conservation law (\ref{cons}) then reduces to the first law of
thermodynamics $\delta\mathcal{F}=0$, where $\mathcal{F}=M-TS$ stands
for the free energy. Hence, the black hole entropy can be expressed in
terms of the surface integral
\begin{equation}
S=-2\pi\int_{0}^{1}ds\left(  \int_{\Sigma_{h}}\hat{\varepsilon}_{\mu\nu}%
\tilde{k}_{\xi}^{\left[  \nu\mu\right]  }d\phi\right)=\frac{\pi b}{4G},
\end{equation}
where $\hat{\varepsilon}_{\mu\nu}$ is the binormal to the bifurcation
surface $\Sigma_{h}$, normalized as
$\hat{\varepsilon}_{\mu\nu}\hat{\varepsilon}%
^{\mu\nu}=-2$. It is not surprising that this result differs from a
quarter of the area of the event horizon since we are not dealing with
general relativity.

\bigskip As a final remark, it is worth pointing out that the results obtained
here can also be generalized to the case of asymptotically locally flat hairy
black holes with angular momentum (\ref{RotatingBH}). Curiously, this black
hole has vanishing angular velocity for the horizon, i.e.,
$\Omega_{+}=0$ so that the first law reduces to
\begin{equation}
dM=TdS-\Omega_{+}dJ=TdS\ .
\end{equation}

\section{Acknowledgments}

The results presented here rely on our preprint \cite{Pucon}. We thank Oscar
Fuentealba, Hernán González, Javier Matulich and Alfredo Pérez for
enlightening discussions. C.T. is a Laurent Houart postdoctoral fellow. R.T.
wishes to thank the Physique théorique et mathématique group of the Université
Libre de Bruxelles, and the International Solvay Institutes for the warm
hospitality. This work is partially funded by the Fondecyt grants N%
${{}^o}$
1130658, 1121031, 11130260, 3140125. The work of G.B. is partially supported
by research grants of the F.R.S.-FNRS and IISN-Belgium as well as the
``Communauté française de Belgique - Actions de Recherche
Concertees\textquotedblright. The work of D.T. is partially supported by the
ERC Advanced Grant ``SyDuGraM'', by a Marina Solvay fellowship, by
FNRS-Belgium (convention FRFC PDR T.1025.14 and convention IISN 4.4514.08) and
by the ``Communauté Française de Belgique'' through the ARC program. The
Centro de Estudios Científicos (CECs) is funded by the Chilean Government
through the Centers of Excellence Base Financing Program of Conicyt.


\begin{thebibliography}{99}                                                                                               %
\bibitem {BTZ}M.~Banados, C.~Teitelboim and J.~Zanelli, ``The Black hole in
three-dimensional space-time, ''Phys.\ Rev.\ Lett.\ \textbf{69},1849 (1992)
{[}arXiv:hep-th/9204099{]}.


\bibitem {BHTZ}M.~Banados, M.~Henneaux, C.~Teitelboim and J.~Zanelli,
``Geometry of the (2+1) black hole,''Phys.\ Rev.\ D \textbf{48}, 1506 (1993)
{[}arXiv:gr-qc/9302012{]}.


\bibitem {BHT0}E.~A.~Bergshoeff, O.~Hohm and P.~K.~Townsend, ``Massive Gravity
in Three Dimensions,''Phys.\ Rev.\ Lett.\ \textbf{102}, 201301 (2009)
{[}arXiv:0901.1766 {[}hep-th{]}{]}.


\bibitem {BHT1}E.~A.~Bergshoeff, O.~Hohm and P.~K.~Townsend, ``More on Massive
3D Gravity,''Phys.\ Rev.\ D \textbf{79}, 124042(2009) {[}arXiv:0905.1259
{[}hep-th{]}{]}.


\bibitem {Deser-Alas}S.~Deser, ``Ghost-free, finite, fourth order D=3 (alas)
gravity, ''Phys.\ Rev.\ Lett.\ \textbf{103},101302 (2009) {[}arXiv:0904.4473
{[}hep-th{]}{]}.


\bibitem {OTT}J.~Oliva, D.~Tempo and R.~Troncoso, \textquotedblleft
Three-dimensional black holes, gravitational solitons, kinks and wormholes for
BHT massive gravity,\textquotedblright\ JHEP\textbf{0907},011 (2009)
{[}arXiv:0905.1545 {[}hep-th{]}{]}.


\bibitem {Joao-Pessoa}J.~Oliva, D.~Tempo and R.~Troncoso, \textquotedblleft
Static spherically symmetric solutions for conformal gravity in three
dimensions,\textquotedblright\ Int.\ J.\ Mod.\ Phys.\ A \textbf{24}, 1588
(2009) {[}arXiv:0905.1510 {[}hep-th{]}{]}.

\bibitem {Blagojevic}  M.~Blagojevi\'c and B.~Cvetkovi\'c,
  ``Conformally flat black holes in Poincar\'e gauge theory,''
  arXiv:1510.00069 [gr-qc].
  
  
\bibitem {Gibbons:2004ai}G.~W.~Gibbons, M.~J.~Perry and C.~N.~Pope, ``The
First Law of Thermodynamics for Kerr-Anti-de Sitter Black Holes,
''Class.\ Quant.\ Grav.\ \textbf{22},1503 (2005) {[}arXiv:hep-th/0408217{]}.


\bibitem {ABS-AS-Null-3D-GR}A.~Ashtekar, J.~Bicak and B.~G.~Schmidt,
``Asymptotic structure of symmetry reduced general
relativity,''Phys.\ Rev.\ D\textbf{55},669 (1997) {[}arXiv:gr-qc/9608042{]}.


\bibitem {GC-Null-Conditions-3D}G.~Barnich and G.~Compere, ``Classical central
extension for asymptotic symmetries at null infinity in three spacetime
dimensions,''Class.\ Quant.\ Grav.\ \textbf{24},F15 (2007) {[}%
arXiv:gr-qc/0610130{]}, Corrigendum-ibid. \textbf{24}, 3139 (2007).


\bibitem {BMSaspect-1}G.~Barnich and C.~Troessaert, ``Aspects of the BMS/CFT
correspondence,'' JHEP \textbf{1005}, 062 (2010) {[}arXiv:1001.1541
{[}hep-th{]}{]}.


\bibitem {Covariant-Barnich-Brandt}G.~Barnich and F.~Brandt, ``Covariant
theory of asymptotic symmetries, conservation laws and central charges,
''Nucl.\ Phys.\ B\textbf{633},3 (2002) {[}arXiv:hep-th/0111246{]}.


\bibitem {GC-Boundary-Charges}G.~Barnich, ``Boundary charges in gauge
theories: Using Stokes theorem in the bulk,
''Class.\ Quant.\ Grav.\ \textbf{20}, 3685(2003) {[}arXiv:hep-th/0301039{]}.


\bibitem {RobinsonTrautman-1}I.~Robinson and A.~Trautman, ``Spherical
Gravitational Waves, ''Phys.\ Rev.\ Lett.\ \textbf{4}, 431 (1960).


\bibitem {RobinsonTrautman-2}I.~Robinson and A.~Trautman, \textquotedblleft
Some spherical gravitational waves in general relativity, \textquotedblright
Proc.\ Roy.\ Soc.\ Lond.\ A \textbf{265}, 463 (1962).


\bibitem {KS-Extended-1}B.~Ett and D.~Kastor, \textquotedblleft An Extended
Kerr-Schild Ansatz,\textquotedblright\ Class.\ Quant.\ Grav.\ \textbf{27},
185024 (2010) {[}arXiv:1002.4378 {[}hep-th{]}{]}.


\bibitem {KS-Extended-2}T.~Málek, ``Extended Kerr-Schild spacetimes: General
properties and some explicit examples,'' Class.\ Quant.\ Grav.\ \textbf{31},
185013 (2014) {[}arXiv:1401.1060 {[}gr-qc{]}{]}.

\bibitem {KC-Anzats-BHT-1}E.~Ayón-Beato, M.~Hassaine and M.~M.~Juárez-Aubry,
\textquotedblleft Towards the uniqueness of Lifshitz black holes and solitons
in New Massive Gravity,\textquotedblright\ Phys.\ Rev.\ D \textbf{90}, no. 4,
044026 (2014) {[}arXiv:1406.1588 {[}hep-th{]}{]}.


\bibitem {KC-Anzats-EGB-1}A.~Anabalon, N.~Deruelle, Y.~Morisawa, J.~Oliva,
M.~Sasaki, D.~Tempo and R.~Troncoso, ``Kerr-Schild ansatz in
Einstein-Gauss-Bonnet gravity: An exact vacuum solution in five dimensions,''
Class.\ Quant.\ Grav.\ \textbf{26}, 065002 (2009) {[}arXiv:0812.3194
{[}hep-th{]}{]}.


\bibitem {KC-Anzats-EGB-2}A.~Anabalon, N.~Deruelle, D.~Tempo and R.~Troncoso,
\textquotedblleft Remarks on the Myers-Perry and Einstein Gauss-Bonnet
Rotating Solutions,\textquotedblright\ Int.\ J.\ Mod.\ Phys.\ D \textbf{20},
639 (2011) {[}arXiv:1009.3030 {[}gr-qc{]}{]}.


\bibitem {KC-Anzats-Lovelock-3}B.~Ett and D.~Kastor, \textquotedblleft
Kerr-Schild Ansatz in Lovelock Gravity,\textquotedblright\ JHEP \textbf{1104},
109 (2011) {[}arXiv:1103.3182 {[}hep-th{]}{]}.

\bibitem{Barnich:2006av}
  G.~Barnich and G.~Compere,
  ``Classical central extension for asymptotic symmetries at null infinity in three spacetime dimensions,''
  Class.\ Quant.\ Grav.\  {\bf 24} (2007) F15
  doi:10.1088/0264-9381/24/5/F01, 10.1088/0264-9381/24/11/C01
  [gr-qc/0610130].

\bibitem{Barnich:2012aw}
  G.~Barnich, A.~Gomberoff and H.~A.~Gonzalez,
  ``The Flat limit of three dimensional asymptotically anti-de Sitter spacetimes,''
  Phys.\ Rev.\ D {\bf 86} (2012) 024020
  doi:10.1103/PhysRevD.86.024020
  [arXiv:1204.3288 [gr-qc]].

\bibitem{Bagchi:2012yk}
  A.~Bagchi, S.~Detournay and D.~Grumiller,
  ``Flat-Space Chiral Gravity,''
  Phys.\ Rev.\ Lett.\  {\bf 109} (2012) 151301
  doi:10.1103/PhysRevLett.109.151301
  [arXiv:1208.1658 [hep-th]].


\bibitem {CosmoEntro-1}G.~Barnich, ``Entropy of three-dimensional
asymptotically flat cosmological solutions, ''JHEP \textbf{1210}, 095 (2012)
{[}arXiv:1208.4371 {[}hep-th{]}{]}.


\bibitem {FlatHolography-1}A.~Bagchi, S.~Detournay, R.~Fareghbal and J.~Simón,
``Holography of 3D Flat Cosmological Horizons,''
Phys.\ Rev.\ Lett.\ \textbf{110}, no. 14, 141302 (2013) {[}arXiv:1208.4372
{[}hep-th{]}{]}.

\bibitem{Bagchi:2013lma}
  A.~Bagchi, S.~Detournay, D.~Grumiller and J.~Simon,
  ``Cosmic Evolution from Phase Transition of Three-Dimensional Flat Space,''
  Phys.\ Rev.\ Lett.\  {\bf 111} (2013) 18,  181301
  doi:10.1103/PhysRevLett.111.181301
  [arXiv:1305.2919 [hep-th]].

\bibitem{Barnich:2013axa}
  G.~Barnich and C.~Troessaert,
  ``Comments on holographic current algebras and asymptotically flat four dimensional spacetimes at null infinity,''
  JHEP {\bf 1311} (2013) 003
  doi:10.1007/JHEP11(2013)003
  [arXiv:1309.0794 [hep-th]].

\bibitem {stromingerBMS-2}A.~Strominger, ``On BMS Invariance of Gravitational
Scattering,'' JHEP \textbf{1407}, 152 (2014) {[}arXiv:1312.2229 {[}%
hep-th{]}{]}.


\bibitem {stromingerBMS-1}T. ~He,V.~Lysov, P.~Mitra and A.~Strominger,
\textquotedblleft BMS supertranslations and Weinberg%
\'{}%
s soft graviton theorem,\textquotedblright\ JHEP \textbf{1505}, 151
(2015){[}arXiv:1401.7026 {[}hep-th{]}{]}.


\bibitem {StromingerBMS-3}F.~Cachazo and A.~Strominger, ``Evidence for a New
Soft Graviton Theorem,'' arXiv:1404.4091 {[}hep-th{]}.

\bibitem{Strominger:2014pwa}
  A.~Strominger and A.~Zhiboedov,
  ``Gravitational Memory, BMS Supertranslations and Soft Theorems,''
  arXiv:1411.5745 [hep-th].

\bibitem{Barnich:2015mui}
  G.~Barnich, H.~A.~Gonzalez, A.~Maloney and B.~Oblak,
  ``One-loop partition function of three-dimensional flat gravity,''
  JHEP {\bf 1504} (2015) 178
  doi:10.1007/JHEP04(2015)178
  [arXiv:1502.06185 [hep-th]].


\bibitem {GOTT}G.~Giribet, J.~Oliva, D.~Tempo and R.~Troncoso, ``Microscopic
entropy of the three-dimensional rotating black hole of BHT massive gravity,''
Phys.\ Rev.\ D \textbf{80}, 124046 (2009) {[}arXiv:0909.2564 {[}hep-th{]}{]}.


\bibitem {PTT}A.~Perez, D.~Tempo and R.~Troncoso, \textquotedblleft
Gravitational solitons, hairy black holes and phase transitions in BHT massive
gravity,\textquotedblright\ JHEP \textbf{1107}, 093 (2011) {[}arXiv:1106.4849
{[}hep-th{]}{]}.


\bibitem {FareghbalHosseini}R.~Fareghbal and S.~M.~Hosseini, ``Holography of
3D Asymptotically Flat Black Holes,'' Phys.\ Rev.\ D \textbf{91}, no. 8,
084025 (2015) {[}arXiv:1412.2569 {[}hep-th{]}{]}.


\bibitem {Bondi-News-1}H.~Bondi, M.~G.~J.~van der Burg and A.~W.~K.~Metzner,
``Gravitational waves in general relativity. 7. Waves from axisymmetric
isolated systems, ''Proc.\ Roy.\ Soc.\ Lond.\ A \textbf{269}, 21 (1962).


\bibitem {Sachs-News-2}R.~K.~Sachs, ``Gravitational waves in general
relativity. VIII. Waves in asymptotically flat space-times,
''Proc.\ Roy.\ Soc.\ Lond.\ A \textbf{270}, 103 (1962).

\bibitem {Wald-Entropy}R.~M.~Wald, ``Black hole entropy is the Noether charge,
''Phys.\ Rev.\ D\textbf{48}, 3427 (1993) {[}arXiv:gr-qc/9307038{]}.

\bibitem{Iyer:1994ys}
  V.~Iyer and R.~M.~Wald,
  Phys.\ Rev.\ D {\bf 50} (1994) 846
  doi:10.1103/PhysRevD.50.846
  [gr-qc/9403028].

\bibitem {Pucon}Glenn Barnich, Cedric Troessaert, David Tempo, Ricardo
Troncoso, ``Asymptotically locally flat spacetimes and black holes in three
dimensions,'' Proceedings of ``XVII Simposio Chileno de Fisica''; Sociedad
Chilena de Fisica, Pucon, Chile, Nov. 2010. Preprints: CECS-PHY-10/13; ULB-TH/10-38.
\end{thebibliography}
\end{document}